\documentclass[english,aip,pop,superscriptaddress,preprint,letterpaper]{revtex4-2}
\usepackage[T1]{fontenc}
\usepackage[utf8]{inputenc}
\usepackage[english]{babel}
\usepackage{graphicx}
\usepackage[dvipsnames]{xcolor}
\usepackage{siunitx,physics}
\usepackage[unicode=true,bookmarks=true,breaklinks=true,backref=false,colorlinks=true,allcolors=Blue]{hyperref}
\usepackage{stickstootext}
\usepackage[stix2]{newtxmath}
\usepackage[capitalize]{cleveref}  % The the cleveref package has to be placed after the hyperref one.
\usepackage{microtype}

\begin{document}

\title{Characterization of mesoscopic turbulent transport events with long-radial-range correlation in DIII-D H-mode plasmas}
\author{R.~Hong}
\email{rohong@ucla.edu}
\affiliation{University of California, Los Angeles, California 90095,~USA}
\author{T.~L.~Rhodes}
\affiliation{University of California, Los Angeles, California 90095,~USA}
% \author{N.~T.~Howard}
% \affiliation{MIT Plasma Science and Fusion Center, Cambridge, MA 02139,~USA}
\author{Y.~Ren}
\affiliation{Princeton Plasma Physics Laboratory, Princeton, New Jersey 08543,~USA}
\author{P.~H.~Diamond}
\affiliation{University of California, San Diego, La Jolla, California 92093,~USA}
\author{X.~Jian}
\affiliation{General Atomics, San Diego, California 92121,~USA}
\author{L.~Zeng}
\affiliation{University of California, Los Angeles, California 90095,~USA}
\author{K.~Barada}
\affiliation{University of California, Los Angeles, California 90095,~USA}
\author{Z.~Yan}
\affiliation{University of Wisconsin, Madison, Wisconsin 53706,~USA}
\author{G.~R.~McKee}
\affiliation{University of Wisconsin, Madison, Wisconsin 53706,~USA}
% \author{the DIII-D team}

\begin{abstract}
A dimensionless collisionality scan has been performed in H-mode plasmas on DIII-D tokamak, with detailed measurements of intermediate-to-high wavenumber turbulence using Doppler backscattering systems.
It is found that the shorter wavelength turbulence develops into spatially asymmetric turbulent structures with a long-radial-range correlation (LRRC) in the mid-radius region of high-collisionality discharges. 
Linear \textsc{cgyro} simulations indicate that the underlying turbulence is likely driven by the electron-temperature-gradient (ETG) mode. 
The LRRC transport events are highly intermittent and show a power spectrum of \(S_{\tilde{n}}(k_\perp) \propto k^{-1}_\perp\) for density fluctuations, which is often associated with self-organized criticality.
The magnitude and the radial scale of those turbulent structures increase significantly when the $E_{r}\times B$ mean flow shearing rate decreases.
The enhanced LRRC transport events appear to be correlated with the degraded energy confinement time.
The emergence of such LRRC transport events may serve as a candidate explanation for the degrading nature of \emph{H}-mode core plasma confinement at high collisionality.
\end{abstract}

\maketitle

\section{Introduction}

A key feature of tokamak plasmas is that the cross-field fluxes of heat and particles are dominated by multiple turbulent processes.\cite{liewerMeasurementsMicroturbulenceTokamaks1985,woottonFluctuationsAnomalousTransport1990,doyleChapterPlasmaConfinement2007}
Due to the inherent complexity of underlying turbulence, the accurate prediction of transport processes and plasma confinement in fusion plasmas is a major challenge for ITER and future fusion reactors.
One of the practical approaches that are used to project the plasma performance of future devices is scaling from present-day machines using parameters that incorporate essential plasma physics and machine conditions.\cite{doyleChapterPlasmaConfinement2007,luceApplicationDimensionlessParameter2008}
In this approach, dimensional analysis and similarity theory are applied to determine the relevant dimensionless quantities that characterize the plasma dynamics.
Several non-dimensional scaling has been obtained for heat transport based on this approach, such as relative gyroradius,\cite{pettyNondimensionalTransportScaling1995,cordeyITERSimulationExperiments1996} collisionality,\cite{greenwaldTransportPhenomenaAlcator1998,pettyScalingHeatTransport1999,zastrowTritiumTransportExperiments2004} and beta scaling.\cite{pettyBetaScalingTransport2004,christiansenScalingConfinement1998}

It has been observed that the normalized energy confinement time shows an inverse dependence on collisionality in \emph{H}-mode plasmas on tokamaks,\cite{luceApplicationDimensionlessParameter2008} i.e., $B\tau_{E} \propto {\nu^{*}}^{-\alpha}$ with $0.3<\alpha<1$ for different devices.
The collisionality, $\nu^{*}\propto\frac{naq}{T^{2}}$, is the collision frequency normalized to the bounce time of a trapped thermal particle, where $n$ is the plasma density, $a$ is the minor radius, \(q\) is the safety factor, and $T$ is the temperature of particle species.
Detailed local power balance analyses confirm that the core electron thermal diffusivity also increases when the collisionality is raised,\cite{pettyScalingHeatTransport1999} indicating that core turbulence plays an essential role in the degradation of energy confinement.

Several theoretical models have been proposed to explain the degraded energy confinement time at higher collisionality in \emph{H}-mode plasmas.
It was suspected that the dominant mode switches from the ion-temperature-gradient (ITG) mode to the resistive ballooning mode (RBM) at higher collisionality.
% \replaced{The RBM can be the dominant mode near the plasma boundary, but typically it is not the primary mode in the core region.}{However, the RBM can become the dominant mode in the region close to the plasma boundary, rather than in the core region.}
The RBM can be the dominant mode at the plasma boundary, but typically it is not the primary mode in the core region.
It is also proposed that the decrease in confinement may be explained by the collisional damping of the zonal flows that regulate turbulence via the shear decorrelation mechanism.
A recent experiment on DIII-D found that the ion-scale long-wavelength turbulence ($k_{\theta}\rho_{s}<1$) has similar amplitude during the collisionality scan, while the intermediate-to-short wavelength turbulence shows higher magnitude at higher collisionality.\cite{mordijckCollisionalityDrivenTurbulent2020}
These observations are inconsistent with the hypothesis that the confinement degradation results from the collisional damping of zonal flows, as the shear decorrelation mechanism should be applicable to both long- and short-wavelength turbulence. 
The microtearing mode\cite{guttenfelderElectromagneticTransportMicrotearing2011} (MTM) and electron-temperature-gradient (ETG) mode\cite{colyerCollisionalityScalingElectron2017} have also been proposed to explain the observed confinement scaling, but experimental evidence of causality is still lacking.
Above all, the underlying mechanism of the degrading nature with collisionality in \emph{H}-mode plasmas remains unclear.
To gain a better understanding of the subject, more in-depth and comprehensive measurements of the core turbulence are required in high collisionality plasmas.

In this study, we carried out a dimensionless collisionality scan in \emph{H}-mode plasmas on the DIII-D tokamak and performed detailed turbulence measurements in the core region.
It is observed that shorter wavelength turbulence develops into spatially asymmetric turbulent structures with a long-radial-range correlation (LRRC) in the mid-radius region of high-collisionality discharges. 
The magnitude and the radial scale of those turbulent structures increase significantly when the $E_{r}\times B$ mean flow shearing rate is reduced below the turbulent scattering rate. 
The enhanced LRRC transport events are accompanied by apparent degradation of normalized energy confinement time.
The emergence of such LRRC transport events may serve as a candidate explanation for the degrading energy confinement time in high-collisionality \emph{H}-mode plasmas.

The rest of this paper is organized as follows.
The experimental setup and diagnostics are described in \cref{sec:expt_setup}.
In \cref{sec:profiles}, experimental results of the equilibrium profiles and power balance analyses are shown.
The characterization of core turbulence in the mid-radius region, particularly the mesoscopic turbulence with long-radial-range correlation, is presented in \cref{sec:turbulence}.
In \cref{sec:scaling}, the relation between the mesoscopic transport events and mean flow shearing rate, as well as the normalized energy confinement time, is given.
A brief discussion and conclusion are shown in \cref{sec:summary}.

\section{Experimental arrangement\label{sec:expt_setup}}

\begin{figure}
    \includegraphics[width=0.4\columnwidth]{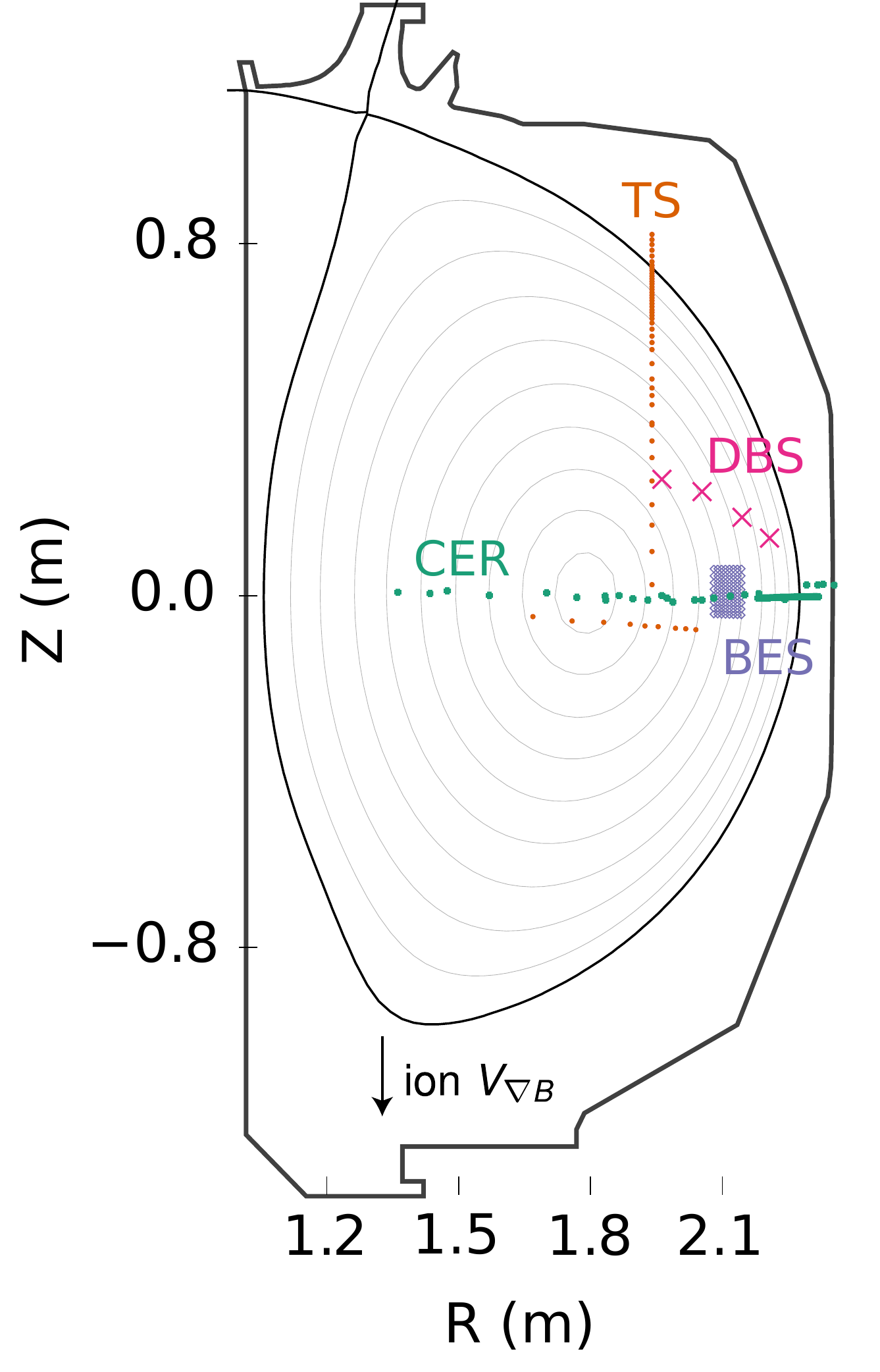}
    \caption{\label{fig:diagnostics}The plasma shaping of a typical discharge used in this study, with an overlay of some major diagnostics.}
\end{figure}

In this study, a dimensionless collisionality scan experiment was performed on the DIII-D tokamak.\cite{luxonDesignRetrospectiveDIIID2002}
Since the standard approach used in dimensionless scan experiments on DIII-D can be found in previous publications,\cite{pettyScalingHeatTransport1999} the experimental setup will be described briefly here.
The inductively-driven hybrid scenario was employed in this study.\cite{pettyHighbetaSteadystateHybrid2015}
The plasma shape is single-null, diverted, with the outer strike point in a closed divertor (\cref{fig:diagnostics}).
The ion $\nabla B$ drift flow is away from the primary divertor.
In this shot-to-shot scan, the collisionality ($\nu_{e}^{*}\sim n_{e}q/T_{e}^{2}$) was varied with fixed plasma density, $n_{e}$.
The toroidal magnetic field $B_{t}$ and the plasma current $I_{p}$ were changed by a factor of $1.6$ with fixed $B_{t}/I_{p}$ or safety factor \(q\).
The electron and ion temperature profiles were well-controlled, i.e., $T_{e,i}\propto B_{t}^{2}$, through varying heating power, to keep the relative gyroradius (\(\rho_*\)) constant.
Here, the neutral beam injection (NBI) of $P_{\mathrm{NBI}}=7.5$ MW provided constant input torque and heating power in each shot, and the electron cyclotron heating (ECH) was also used in some shots to control the electron temperature profile.
The line-averaged plasma density was $\bar{n}_{e}\approx\SI{4e19}{\per\cubic\meter}$ in these discharges.
The corresponding Greenwald fraction was raised from about 0.5 to 0.9.
As a result, the collisionality varied as $\nu_{e}^{*} \propto B_{t}^{-4}$, i.e., by a factor of 7.
Some transport-relevant dimensionless parameters were also well-matched in the mid-radius region in these shots, such as $\rho_{*}$ (ratio of ion gyro-radius to plasma minor radius $\rho_{i}/a$), $\beta_\mathrm{N}=\frac{\beta}{I/aB_{t}}$ (normalized ratio of plasma to magnetic pressure), $q$ (safety factor), $T_{e}/T_{i}$ (electron-to-ion temperature ratio), $\kappa$ (elongation), $\delta$ (triangularity), etc.

The Thomson scattering diagnostic\cite{eldonInitialResultsHigh2012} provided the electron density and temperature profiles.
The charge exchange recombination (CER) diagnostic\cite{chrystalImprovedEdgeCharge2016} provided profiles of carbon ion density, temperature, and rotation velocities.
The radial force balance equation is employed to relate the carbon ion rotation velocities and pressure gradient to the mean $E_{r} \times B$ shear flow.\cite{groebnerRoleEdgeElectric1990}
The plasma current profile was found from an axisymmetric reconstruction of the magnetic equilibrium using the external magnetic measurements, the kinetic profile data, and the magnetic pitch angle measured using the motional Stark effect (MSE).\cite{wroblewskiPolarimetryMotionalStark1992,laoEquilibriumAnalysisCurrent1990}
Part of the data analysis was performed using the OMFIT integrated modeling framework. \cite{meneghiniIntegratedModelingApplications2015} 
The multi-channel Doppler backscattering (DBS) diagnostics,\cite{peeblesNovelMultichannelCombfrequency2010} were utilized to detect the electron density fluctuations at the wavenumber range of $5<k_{\perp}<\SI{10}{\per\cm}$ ($1<k_{\bot}\rho_{s}<4$ where $\rho_{s}$ is the ion gyro-radius with sound speed) in the mid-radius region of the midplane ($0.35<\rho<0.8$ with $\rho$ the normalized minor radius).
The beam emission spectroscopy (BES) diagnostic\cite{mckeeWidefieldTurbulenceImaging2010} was used to provide 2-dimensional long-wavelength ($k_\bot<\SI{3}{\per\cm}$) density turbulence structures at the midplane.

In these shots, the large-scale magnetohydrodynamics (MHD) instabilities were mitigated and of similar amplitude ($\tilde{B}_{\theta}<\SI{0.1}{\milli\tesla}$), which eliminates confounding effects on turbulence dynamics from the background MHD events.
It is worth noting that the pedestal structures and edge localized modes (ELMs) characteristics changed dramatically during the collisionality scan.
To avoid confusion, the analysis presented in this study is focused on the profiles and fluctuations between ELMs in the mid-radius region.

\section{Profiles evolution and power balance analysis\label{sec:profiles}}

\begin{figure}
    \includegraphics[width=0.6\columnwidth]{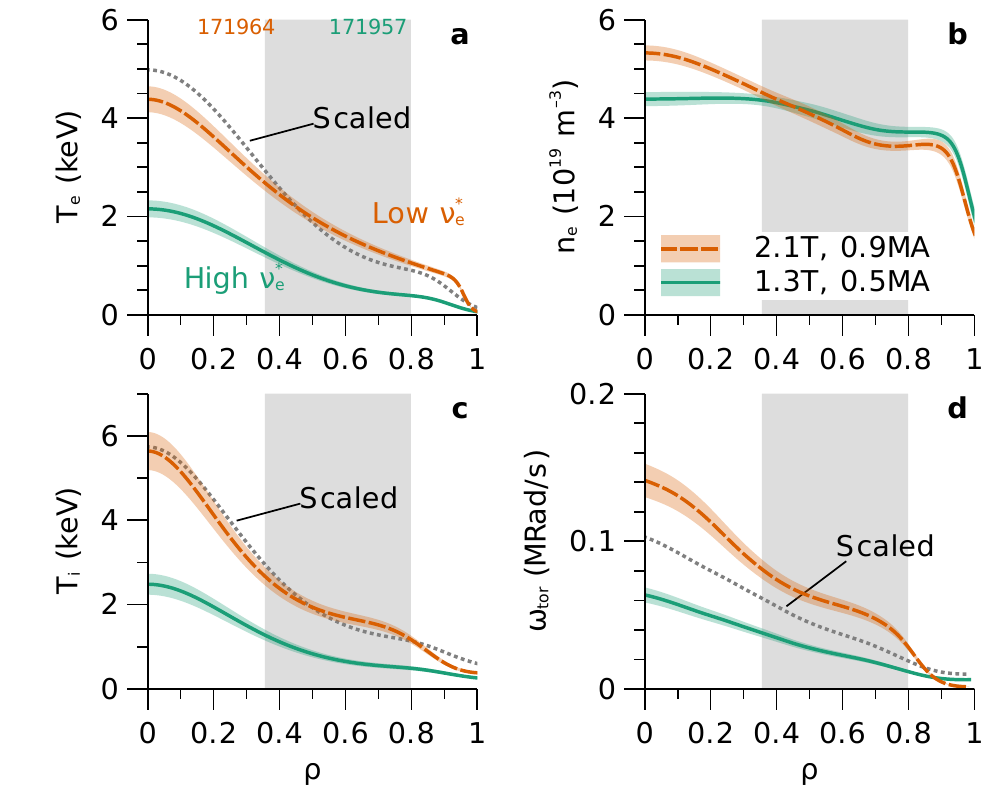}
    \caption{\label{fig:profiles}Radial profiles of (a) electron temperature, (b) electron density, (c) ion temperature, and (d) carbon ion toroidal rotation. The high-\(\nu^*\) profiles are in green at \(B_t=\SI{1.3}{\tesla}\); the low-\(\nu^*\) profiles are in orange at \(B_t=\SI{2.1}{\tesla}\). The dotted curves in (a) and (c) represent the \SI{1.3}{\tesla} profiles scaled to \SI{2.1}{\tesla} using the relation, $T_{e,i}\propto B_t^2$, which is required to keep the Lamor radii constant. The dotted curve in (d) represents the \SI{1.3}{\tesla} profiles scaled to \SI{2.1}{\tesla} with the same Mach number. The shaded area indicates the region of interest in this study.}
\end{figure}

The radial profiles of temperatures, density, and collisionality for discharges during the collisionality scan are illustrated in \cref{fig:profiles}.
The black dotted curves in \cref{fig:profiles} indicate the temperature profiles that are scaled according to $T_{e,i}\propto B_t^2$ from the low-\(B_t\) case for these discharges.
It can be seen that the temperature profile is well controlled, and thus the gyro-radius is held nearly constant in these charges during the scan as $\rho_{e,i}^2 \propto T_{e,i}/B_t^2$.
The line-averaged plasma density is held nearly constant in these discharges, i.e., $\bar{n}_e\approx\SI{4e19}{\per\cubic\meter}$, although the density profile is more peaked in low collisionality shots as shown in \cref{fig:profiles}(b).
Also shown are radial profiles of the ion toroidal rotation.
The carbon ion toroidal rotation and its Mach number drop when the collisionality is raised (\cref{fig:profiles}(d)).
The dotted curve in \cref{fig:profiles}(d) presents the ion toroidal rotation profile scaled with the same Mach number for these discharges.

\begin{figure}
    \includegraphics[width=0.6\columnwidth]{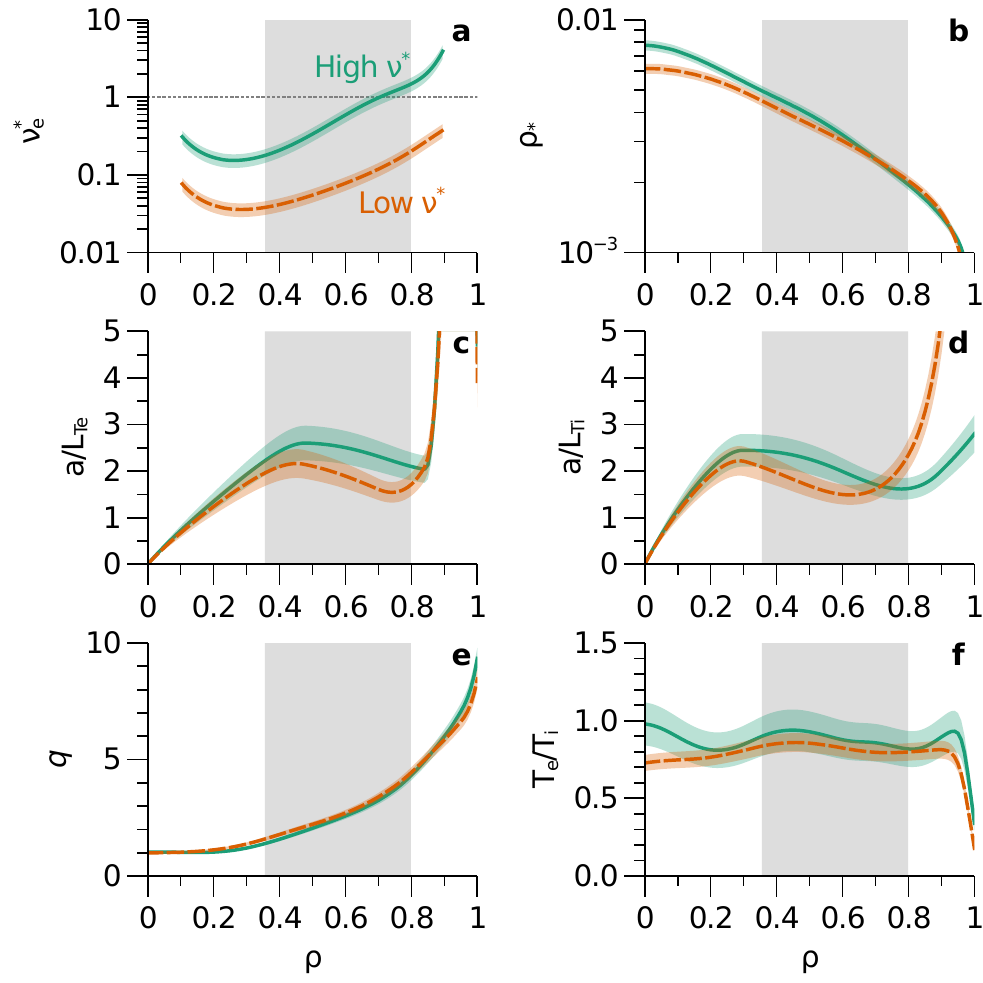}
    \caption{\label{fig:profiles_nondim}Radial profiles of (a) electron collisionality, (b) relative ion gyro-radius, (c) dimensionless electron temperature gradient scale length, (d) dimensionless ion temperature gradient scale length, (e) safety factor, and (f) ratio of electron-to-ion temperature for two discharges with different collisionality. The high-\(\nu^*\) profiles are in green at \(B_t=\SI{1.3}{\tesla}\); the low-\(\nu^*\) profiles are in orange at \(B_t=\SI{2.1}{\tesla}\). The shaded area indicates the region of interest in this study.}
\end{figure}

A comparison of the radial profiles of the dimensionless quantities relevant to turbulent transport is shown in \cref{fig:profiles_nondim}.
The dimensionless electron collisionality, $\nu_{e}^{*}$, is increased by a factor of 7 in the mid-radius region (\cref{fig:profiles_nondim}(a)) in this study.
It also confirms that the local values of $\rho_*$, $T_e/T_i$, $q$, etc., are kept constant as the collisionality is varied by a factor of 7.
It is also worth noting that the dimensionless temperature gradient scale lengths, i.e., $a/L_{T_{e}}$ and $a/L_{T_{i}}$ ($a$ is the plasma minor radius and $1/L_{T_{e,i}}=-\grad_r \ln T_{e,i}$ are the temperature gradient scale lengths), are similar in the mid-radius region during the collisionality scan.

\begin{figure}
	\includegraphics[width=0.6\columnwidth]{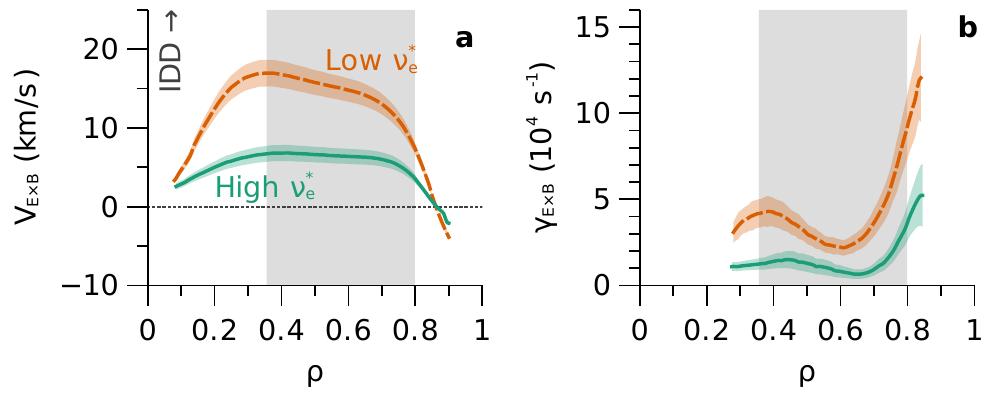}
	\caption{\label{fig:profiles_ExB}Radial profiles of (a) mean $E_r \times B$ shear flow and (b) its radial shearing rate $\gamma_{E\times B} = |\partial_r V_{E\times B}|$. Positive velocity in (a) is the ion diamagnetic drift direction. The shaded area indicates the region of interest in this study.}
\end{figure}

A comparison of the mean $E_{r}\times B$ shear layer has also been performed.
As mentioned in \cref{sec:expt_setup}, the radial electric field, $E_r$, can be inferred using the ion force balance equation, i.e., $ E_r = \frac{\nabla p_i}{n_i Z_i e} - V_{i,\phi} B_\theta + V_{i, \theta} B_\phi$.
Here, $ n_i $ is the density of carbon impurity, $ Z_i e $ is the charge of carbon impurity, $ p_i=n_i T_i $ is the pressure, $ T_i $ is the temperature, $ V_{i,\phi} $ is the toroidal rotation speed, $ V_{i,\theta} $ is the poloidal rotation speed, $ B_\phi $ is the toroidal component of the magnetic field, and $ B_\theta $ is the poloidal component of that field.
In these discharges, the dominant contribution to $E_r$ is the toroidal rotation driven by the NBI torque input, i.e., $ V_{i,\phi} B_\theta $.
As illustrated in \cref{fig:profiles}(d), the ion toroidal rotation drops substantially at higher collisionality.
As a result, the mean $E_{r}\times B$ shear flow and its shearing rate, $ \gamma_{E\times B}=\abs{\nabla_r V_{E\times B}} $, reduces in the core region as the collisionality is raised (\cref{fig:profiles_ExB}).
It will be shown later that the variation in the flow shearing rate is highly related to the changes in turbulent transport processes.

\begin{figure}
    \includegraphics[width=0.6\columnwidth]{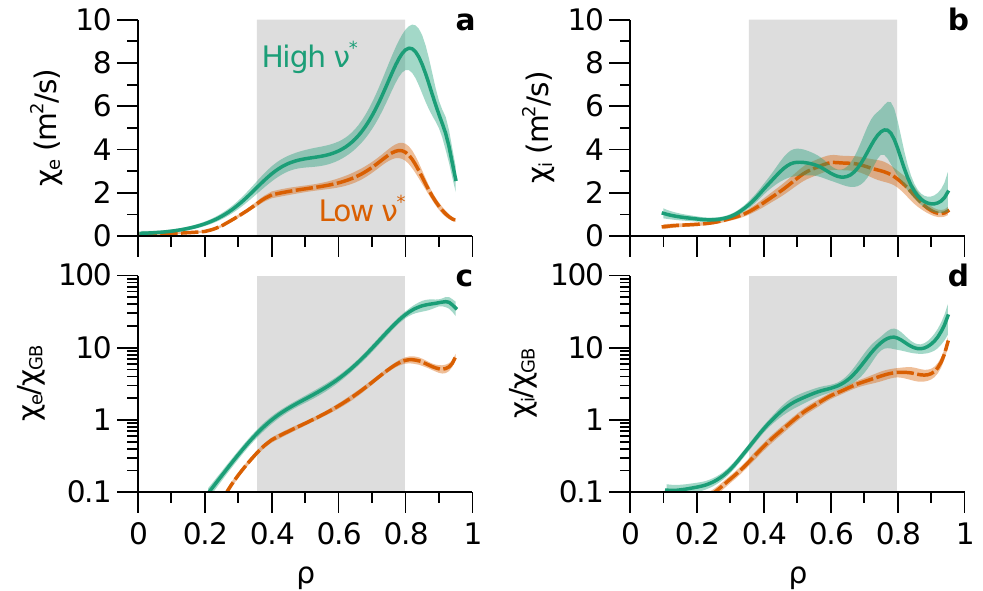}
    \caption{\label{fig:profiles_diffusivity}Radial profiles of (a) electron thermal diffusivity, (b) ion thermal diffusivity, (c) dimensionless electron thermal diffusivity in the gyro-Bohm unit, and (d) dimensionless ion thermal diffusivity in the gyro-Bohm unit. The shaded area indicates the region of interest in this study.}
\end{figure}

The thermal diffusivities can be determined from a power balance analysis of the energy transport.
In this study, the radial power balance equation is solved by the \textsc{transp} transport code,\cite{pankinTokamakMonteCarlo2004} which utilizes the measured profiles of the electron density, electron and ion temperatures, ion rotation, effective ion charge, radiated power, and external power input.
Using the thermal flux calculated by the \textsc{transp} code, the thermal diffusivity is then determined from $ \chi_{e,i} = -\frac{Q_{e,i}}{n\nabla T_{e,i}}$, which presumes purely diffusive thermal fluxes.
It can be seen in \cref{fig:profiles_diffusivity} that the core electron thermal diffusivity is larger for the high collisionality case, while the ion thermal diffusivity is similar during the scan.
This local transport analysis confirms the collisionality dependence observed in the global confinement time for H-mode plasmas.\cite{pettyScalingHeatTransport1999}
These findings also indicate that the electron turbulent transport in the core is substantially enhanced in high collisionality and low flow shear shots.

\section{Characterization of mesoscopic turbulence\label{sec:turbulence}}

\begin{figure}
    \includegraphics[width=0.85\textwidth]{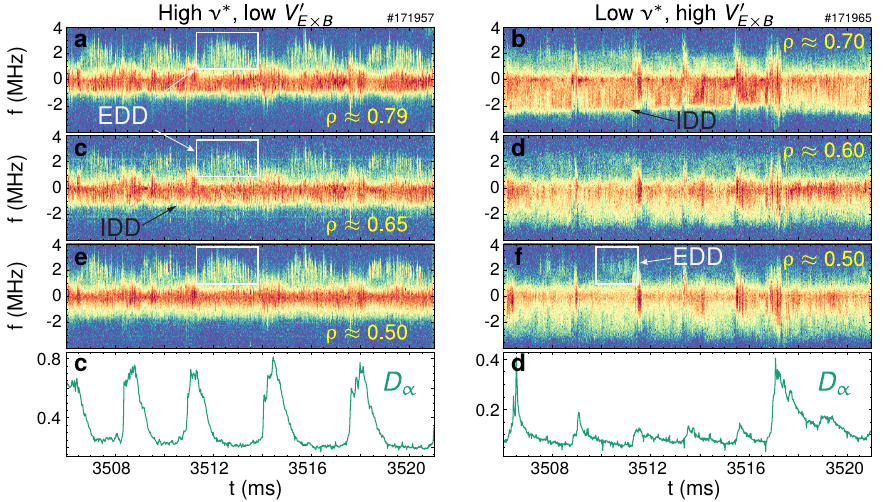}
    \caption{\label{fig:spectrogram}Spectrograms of density fluctuations measured by DBS systems at different radial locations in (left column) high collisionality weak flow shear shots and (right column) low collisionality, strong flow shear discharges. The bottom row shows the time traces of the $ D_\alpha $ emission intensity (in arbitrary units), indicating the ELMs dynamics. The mode moving in the electron diamagnetic drift (EDD) direction is marked by white labels, while the mode propagating in the ion diamagnetic drift (IDD) direction is marked by black labels.}
\end{figure}

With different levels of collisionality and $E_{r}\times B$ shear flows, density fluctuations exhibit distinct spectral and temporal characteristics.
In high collisionality discharges, with a weak $E_{r}\times B$ shear flow, two different modes can be detected by the DBS diagnostics between ELMs (left column in \cref{fig:spectrogram}).
Here, the ELM evolution is indicated by the $D_{\alpha}$ emission intensity.
One mode has a negative Doppler frequency shift corresponding to ion diamagnetic drift (IDD) direction in the lab frame (IDD mode is marked by black labels in \cref{fig:spectrogram}); the other mode has a positive Doppler frequency shift around \SI{2}{\mega\hertz} propagating in the electron diamagnetic drift (EDD) direction in the lab frame (EDD mode is marked by white labels in \cref{fig:spectrogram}).
In low collisionality discharges, with strong $E_{r}\times B$ shear flow, the IDD mode with $f<0$ can still be observed with a larger Doppler shift, which is in agreement with larger background mean flow.
However, the EDD mode around $f\approx\SI{2}{\mega\hertz}$ is only occasionally observed (right column in \cref{fig:spectrogram}), and its amplitude is also significantly reduced in the strong flow, low collisionality case.

\begin{figure}
    \includegraphics[width=0.6\linewidth]{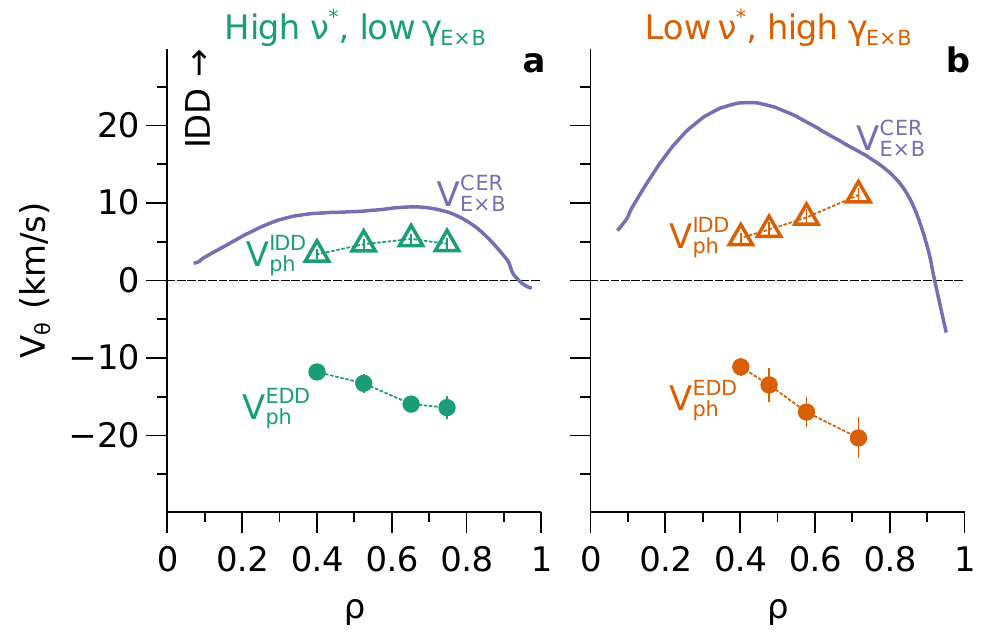}
    \caption{\label{fig:phase_vel}Poloidal phase velocities of IDD and EDD modes measured by DBS are plotted against the mean flow velocity measured by CER in (a) high collisionality and (b) low collisionality shots. The positive poloidal velocity direction represents the ion diamagnetic drift direction.}
\end{figure}

It is worth mentioning that the scales of both IDD and EDD modes are in the sub-ion-gyroradius regime (\(k_{\theta}\rho_{s}>1\)).
By comparing their phase velocities in the lab frame against the mean flow velocity, we can obtain their phase velocities and propagation directions in the plasma rest frame (\cref{fig:phase_vel}).
The mean flow velocity in this study is determined from mean $ E_r \times B $ shear flow measured by the CER systems (purule curves in \cref{fig:phase_vel}).
The phase velocities can be determined from the Doppler shifts, which are measured by DBS diagnostics in this study, together with their corresponding wavenumbers, i.e., $ V_\theta = \frac{\omega_\mathrm{D}}{k_\theta}$.
Here, the wavenumbers of turbulence are calculated using the 3D ray-tracing code, \textsc{genray}.
It can be seen that both the IDD and EDD modes propagate in the electron diamagnetic drift direction in the plasma frame.
Moreover, the phase velocity of the IDD mode is much less than that of the EDD mode in the plasma rest frame.
Hence, in the rest frame of plasmas, the IDD mode is a slow mode, while the EDD mode is a fast mode.

\begin{figure}
    \includegraphics[width=0.6\linewidth]{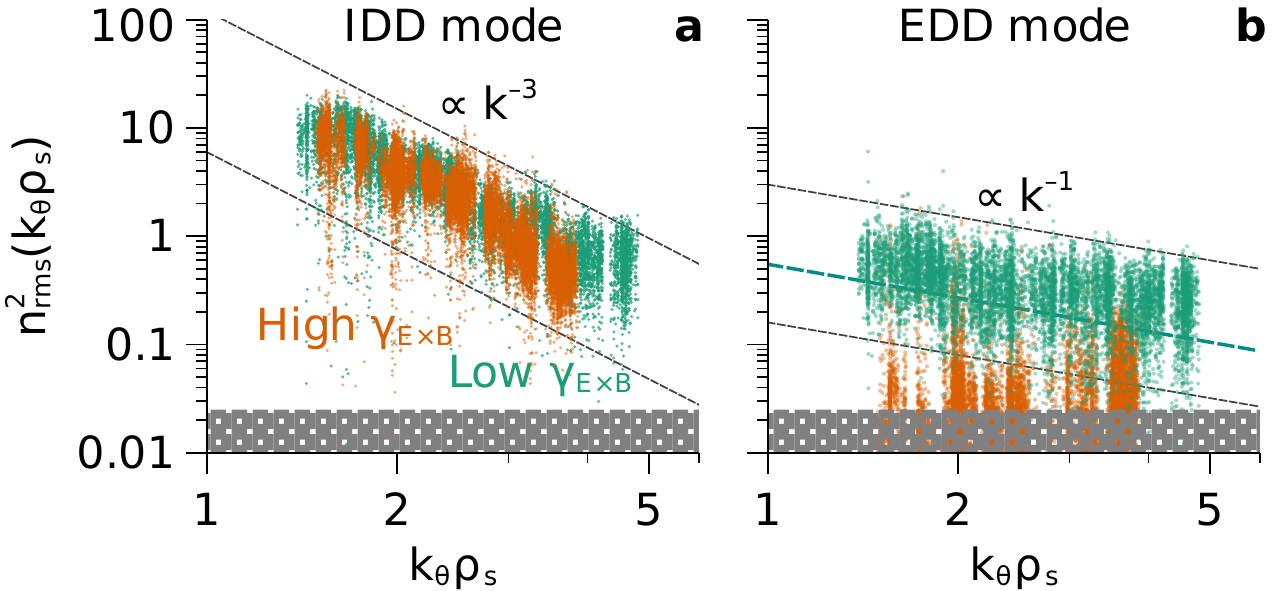}
    \caption{\label{fig:k_spectra}Wavenumber power spectra of (a) IDD mode and (b) EDD mode in both high and low collisionality discharges. The shaded areas indicate the noise level.}
\end{figure}

The wavenumber power spectra of electron density fluctuations, $ S(k_\perp) $, are illustrated in \cref{fig:k_spectra}.
It is found that the wavenumber power spectra of the IDD mode obey a power-law of $S(k_{\perp})\propto k_{\perp}^{-3}$, which is a typical scaling for 2-dimensional turbulence in toroidal magnetic fusion devices.
On the other hand, the EDD mode shows a power-law of $S(k_{\perp})\propto k_{\perp}^{-1}$, indicative of its avalanching-like behavior that is commonly associated with self-organized criticality (SOC).\cite{diamondDynamicsTurbulentTransport1995,bakSelforganizedCriticalityExplanation1987}
Moreover, the magnitudes of the IDD mode are similar during the scan, while the magnitude of the EDD mode is substantially larger in the weak $E_{r}\times B$ shear, high collisionality discharges.
These findings thus suggest that although the IDD mode has a larger amplitude, it is \emph{not} likely responsible for the degradation of the plasma confinement during the scan.

\begin{figure}
    \includegraphics[width=0.6\linewidth]{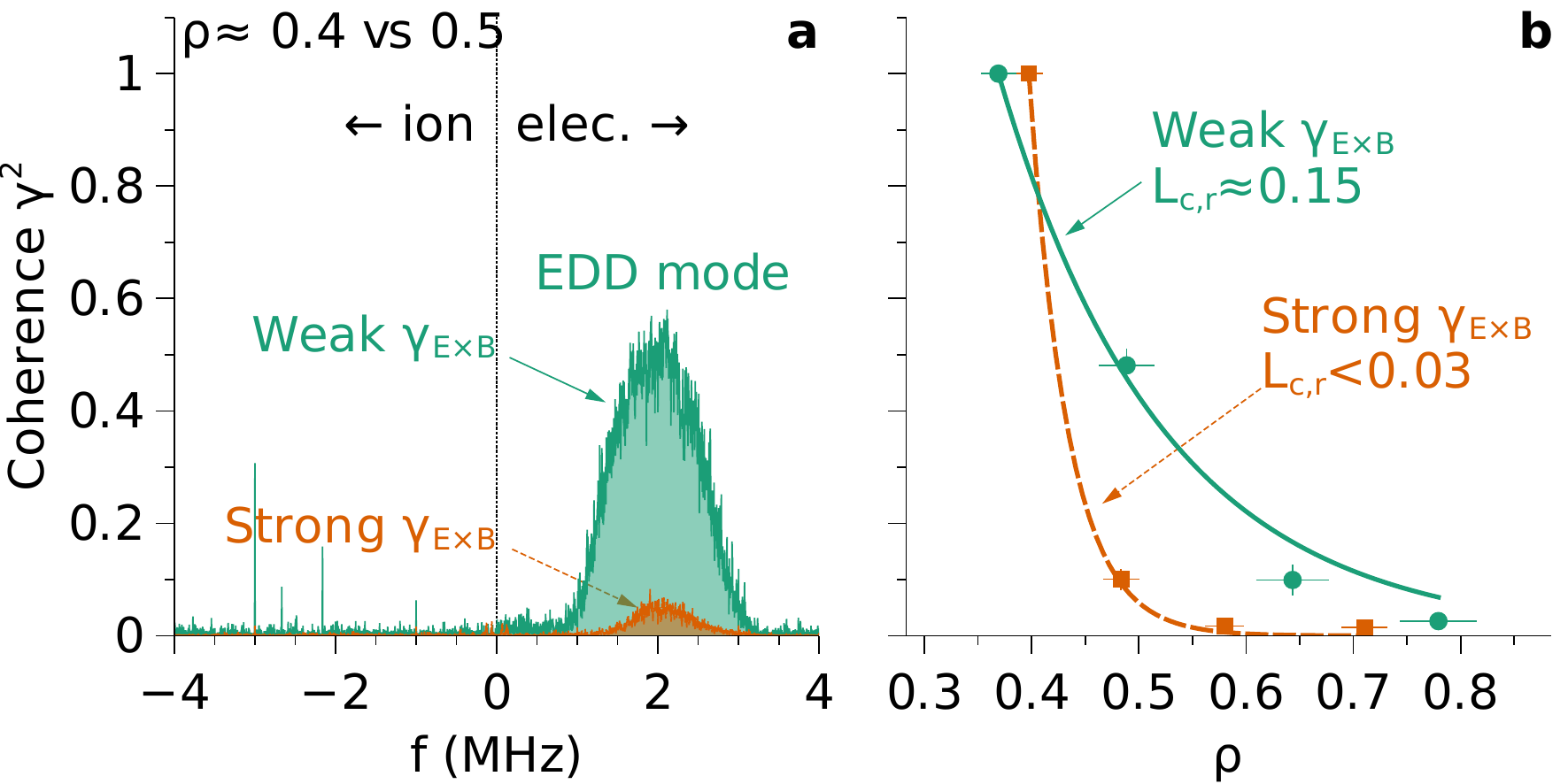}
    \caption{\label{fig:coherence}(a) Frequency-resolved coherence between radially separated density fluctuations in high and low flow shear cases; (b) radial profiles of the peak cross-coherence for the LRRC turbulence, where the channel at $\rho_{0}\approx0.4$ is chosen as the reference. The curves represent the exponential fits, i.e., $\gamma^{2}(\rho)=\exp(-\frac{\rho-\rho_{0}}{L_{c,r}})$, where $L_{c,r}$ represents the radial correlation length in terms of the minor radius.}
\end{figure}

The EDD mode is of particular interest in the study as it exhibits spatial asymmetry and long-radial-range correlation (LRRC), i.e., an LRRC mode.
The frequency-resolved coherence $\gamma^{2}(f)$ has been calculated between electron density fluctuations at different radial locations. 
As shown in \cref{fig:coherence}(a), the fluctuations at $f<\SI{0}{\mega\hertz}$ do not show any clear coherence between signals at $\rho\approx0.4$ and 0.5, while the LRRC mode at $f\approx\SI{2}{\mega\hertz}$ shows substantial coherence with lower mean $E_{r}\times B$ shear flow.
Using the innermost channel at $\rho\approx0.4$ as the reference, we can obtain the radial profiles of the peak coherence for underlying turbulence of the LRRC mode (\cref{fig:coherence}(b)).
The radial correlation length can be determined via the e-folding length of the exponential fits, i.e., $\gamma^{2}(\rho)=\gamma^{2}(\rho)=\exp(-\frac{\rho-\rho_{0}}{L_{c,r}})$, where $L_{c,r}$ represents the e-folding length in terms of the normalized minor radius. 
When converted to physical units, the radial correlation length of the LRRC mode, $L_{c,r}$, is no more than 2 cm (plasma minor radius $a\approx61$ cm) with strong $E_{r}\times B$ shear flow, while it increases to about 10 cm with reduced $E_{r}\times B$ shear flow in high collisionality plasmas.
such long radial correlation length indicates a radially elongated structure of the LRRC mode in low flow shear shots, i.e., $k_{r}\ll k_{\theta}$ with $k_{r}\rho_{s}=0.1-0.3$ and $k_{\theta}\rho_{s}=1-4$.

\begin{figure}
    \includegraphics[width=0.7\linewidth]{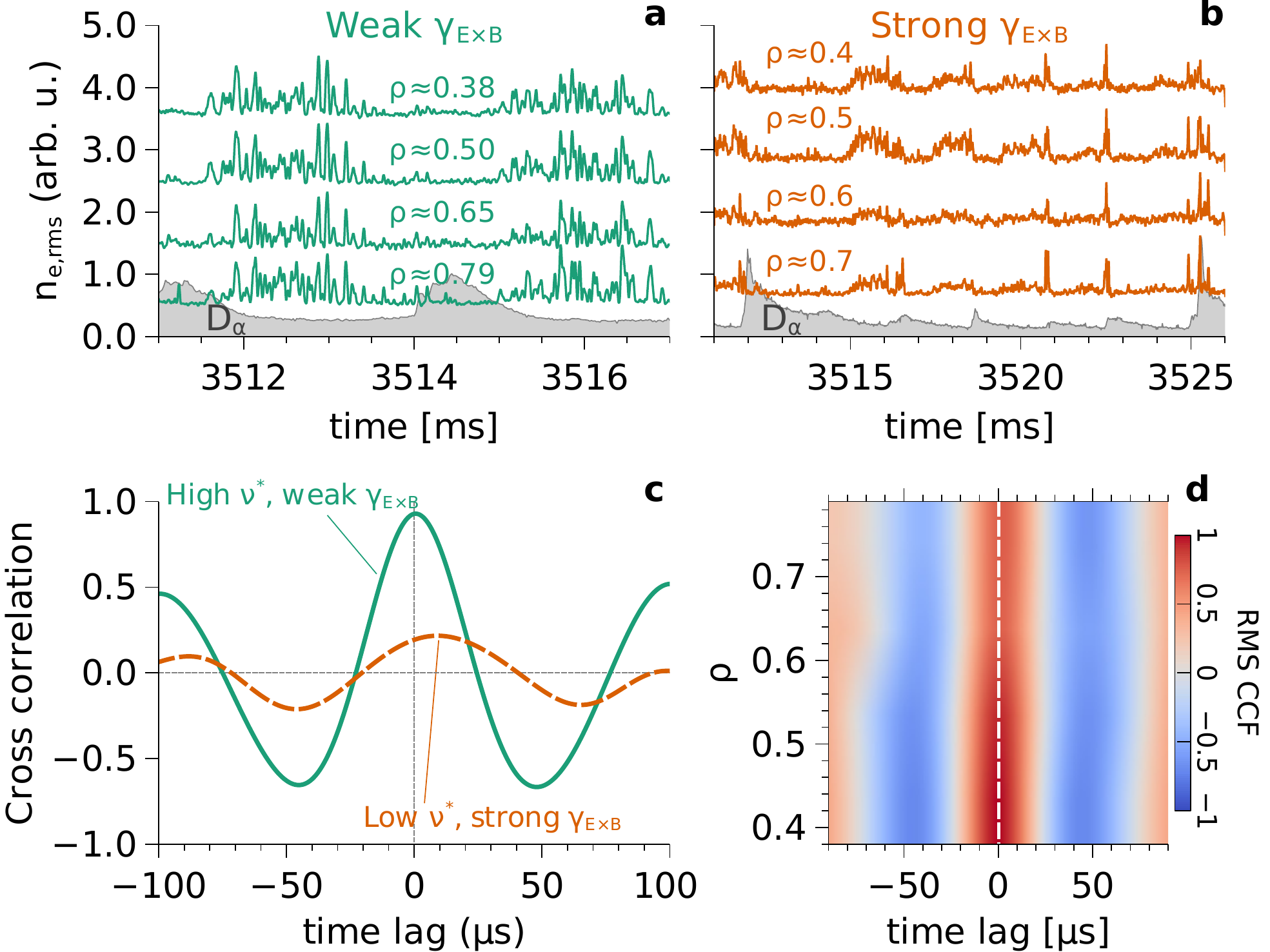}
    \caption{\label{fig:rms_ccf} Time traces of the root-mean-square (RMS) of local density fluctuations (with offsets) corresponding to the LRRC mode at different radial locations in (a) strong and (b) weak flow shear cases, respectively. The $D_{\alpha}$ intensities (gray curves) indicate the ELM dynamics. (c) Cross-correlation of LRRC mode's RMS levels of the neighbor channels ($\rho\approx 0.4$ vs 0.5 in both strong and weak flow shear cases). (d) The contour of the cross-correlation function of the LRRC mode's RMS levels for each DBS channel in high $\nu^*$ and weak $\gamma_{E\times B}$ case, with the innermost channel as the reference.}
\end{figure}

The changes in cross-correlation coefficients can also be identified in the envelopes of the LRRC mode across a wide radial range at low collisionality and weak flow shear.
With weak flow shear, the envelopes, or root-mean-square (RMS) levels, of density fluctuations corresponding to the LRRC mode at different radial locations are well aligned in time (\cref{fig:rms_ccf}(a)).
The time traces of envelopes in strong flow shear are also plotted for comparison (\cref{fig:rms_ccf}(b)).
The cross-correlation coefficient between the envelopes at $\rho\approx0.4$ and 0.5 approaches the value of one with weak flow shear, while the one in strong flow shear case is insignificant (\cref{fig:rms_ccf}(c)). 
By plotting the cross-correlation of the envelopes as a function of the radial location and the time lag, for the weak flow case (\cref{fig:rms_ccf}(d)), with the innermost channel as the reference, one can see that the envelope of the LRRC mode spans a broad radial range of spatial scales ($\rho_{i}\ll\Delta_{r}^{\mathrm{env}}\lesssim a$) with a negligible time delay. 
Such a wide radial scale of the envelopes indicates the emergence of streamer-like transport events in the weak flow shear discharges.

\begin{figure}
    \includegraphics[width=0.45\linewidth]{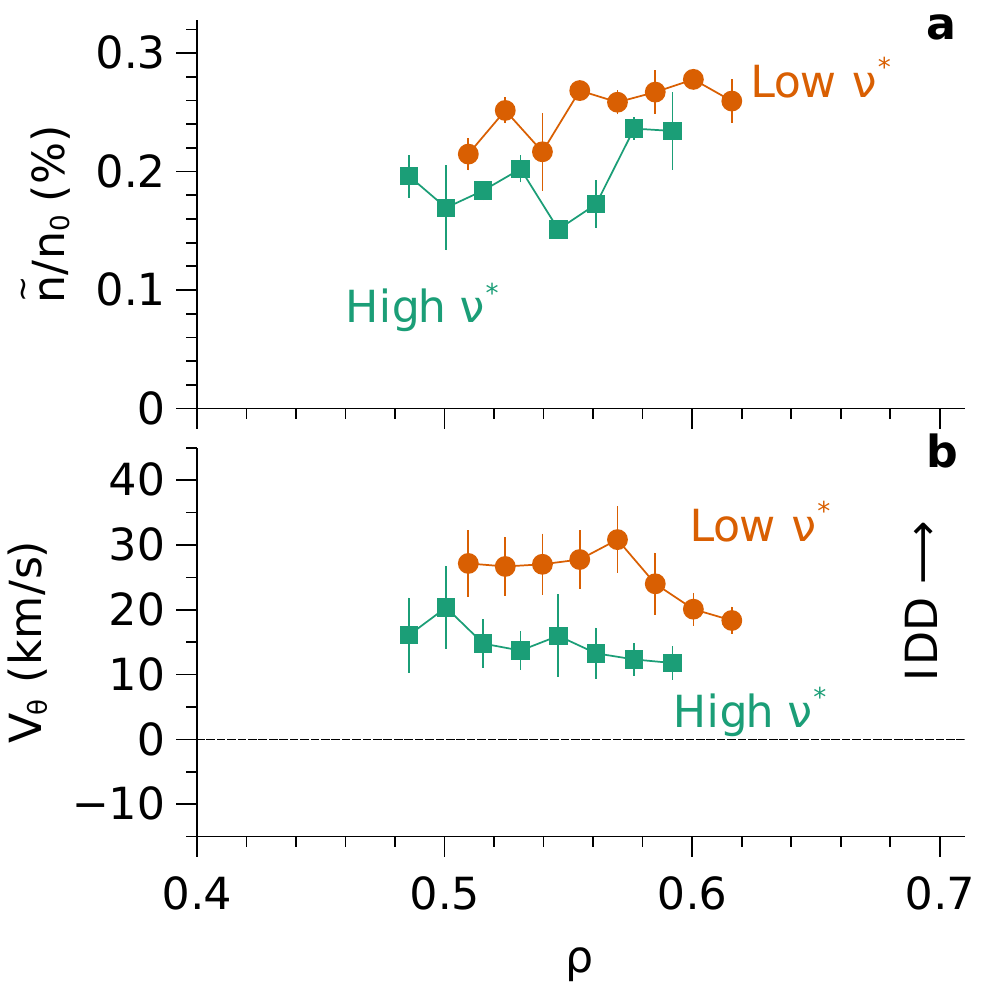}
    \caption{\label{fig:bes_fluct} Radial profiles of (a) normalized amplitude and (b) poloidal phase velocity of low-$k_\theta$ density fluctuations measured by BES diagnostics. The positive phase velocity is in the ion diamagnetic drift direction.}
\end{figure}

In addition to the characterization of turbulence in the sub-ion-gyroradius scale, ion-scale long-wavelength turbulence has also been measured by the beam emission spectroscopy system (BES) in this study.
As shown in \cref{fig:bes_fluct}(a), the normalized magnitude of the ion-scale density fluctuations ($k_\theta\rho_s<1$) is similar or slightly decreases when the collisionality is raised and the mean flow shearing rate is reduced.
Although not shown here, the absolute amplitude of the ion-scale density fluctuations is also similar for those cases, since the variation of the local equilibrium density is less than \(5\%\) during the collisionality scan.
The slightly reduced low-$k_{\theta}$ fluctuations are consistent with previously reported measurements on DIII-D.\cite{mordijckCollisionalityDrivenTurbulent2020}
These findings indicate that the low-$k_\theta$ ion-scale turbulence is unlikely responsible for the changes in the plasma confinement during the collisionality scan.

These ion-scale density fluctuations are propagating in the ion diamagnetic drift direction (\cref{fig:bes_fluct}(b)).
The phase velocity decreases from \SI{30}{\kilo\meter\per\second} to \SI{15}{\kilo\meter\per\second} when the collisionality is raised by a factor of 7.
The decreased phase velocity is consistent with the reduced mean shear layer at higher collisionality.
It is worth mentioning that the amplitude of ion-scale turbulence seems to be independent of the changes in the mean shear layer.
A candidate explanation is that the ion-scale turbulence is mitigated with the presence of the strong toroidal Alfven eigenmodes in high-\(\nu^*\) and low-\(B_t\) shots, as suggested by recent work.\cite{citrinOverviewTokamakTurbulence2023}
This would be a topic for future studies.

\section{Interplay between mean shear flow and mesoscopic turbulence\label{sec:scaling}}

\begin{figure}
  \includegraphics[width=0.7\linewidth]{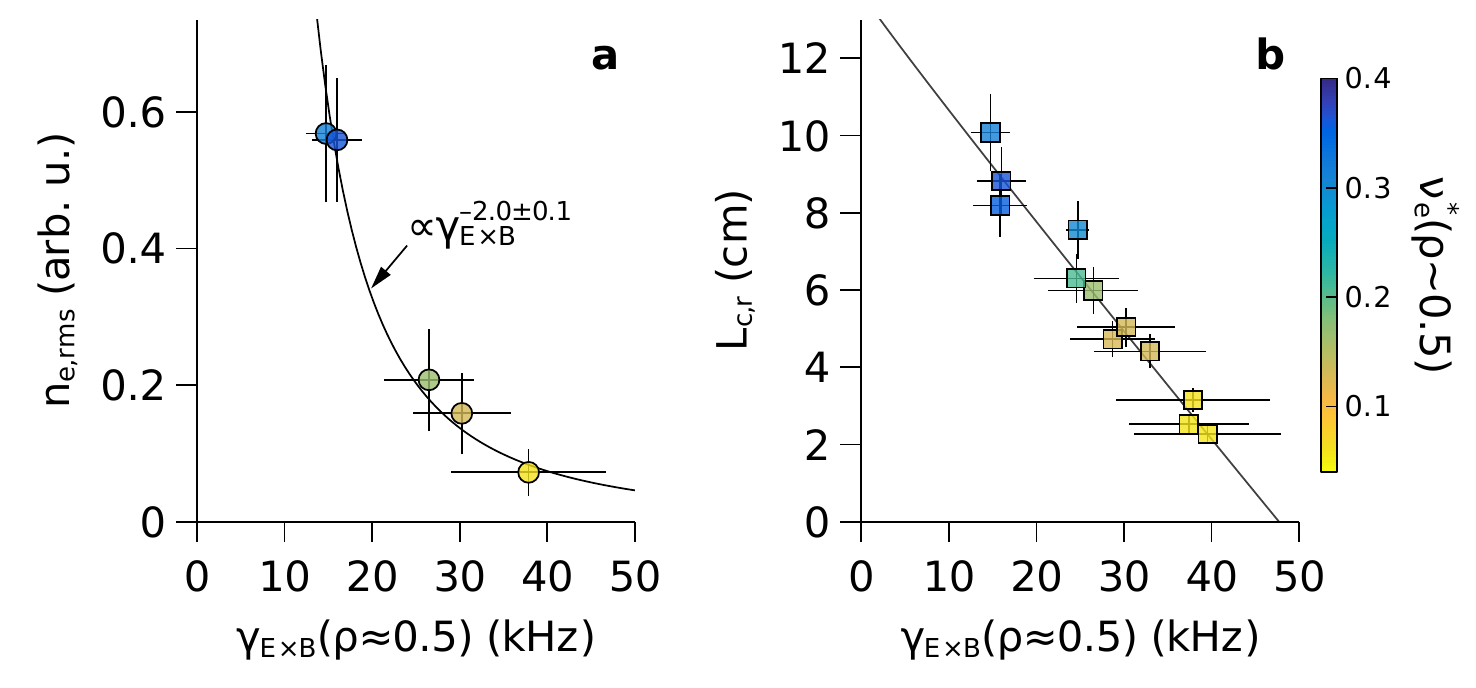}
  \caption{\label{fig:shear_scaling} The amplitude (a) and radial correlation length (b) of underlying fluctuations ($k_{\theta}\rho_{s}\approx3$) of LRRC transport events are plotted against the local mean flow shearing rate at $\rho\approx0.5$. The local collisionality is color-coded with darker markers corresponding to the higher collisionality shots.}
\end{figure}

The observations described in the previous section indicate that the mean shear layer may play an important role in the development of mesoscopic turbulent transport events.
To gain further insight into this relationship, we plot the amplitude and radial scale length of the underlying fluctuations ($k_{\theta}\rho_{s}\approx3$) of LRRC transport events against the mean flow shearing rate (\cref{fig:shear_scaling}).
The local collisionality is color-coded with darker markers corresponding to the higher collisionality shots.
As can be seen in \cref{fig:shear_scaling}(a), the amplitude of the LRRC modes decreases when the local mean flow shearing is raised, i.e., $\tilde{n}_{e,\mathrm{LRRC}}\propto\gamma_{E \times B}^{-2.0\pm0.1}$.
The radial correlation length, on the other hand, is linearly anti-correlated with the mean flow shearing rate (\cref{fig:shear_scaling}(b)).
It is also worth mentioning that the amplitude of the LRRC transport events surges when the mean flow shearing rate is below the turbulent scattering rate, i.e., $\gamma_{E\times B}<\tau_{c}^{-1}$.
Here, the turbulent scattering rate is estimated using the inverse of the auto-correlation time of the LRRC mode in the weak shear flow case ($\tau_{c}^{-1}\approx20-\SI{30}{\kilo\hertz}$).
This finding indicates that the mean shear layer impacts the development of the LRRC turbulent structures, which is consistent with the shear decorrelation mechanism.

\begin{figure}
  \includegraphics[width=0.7\linewidth]{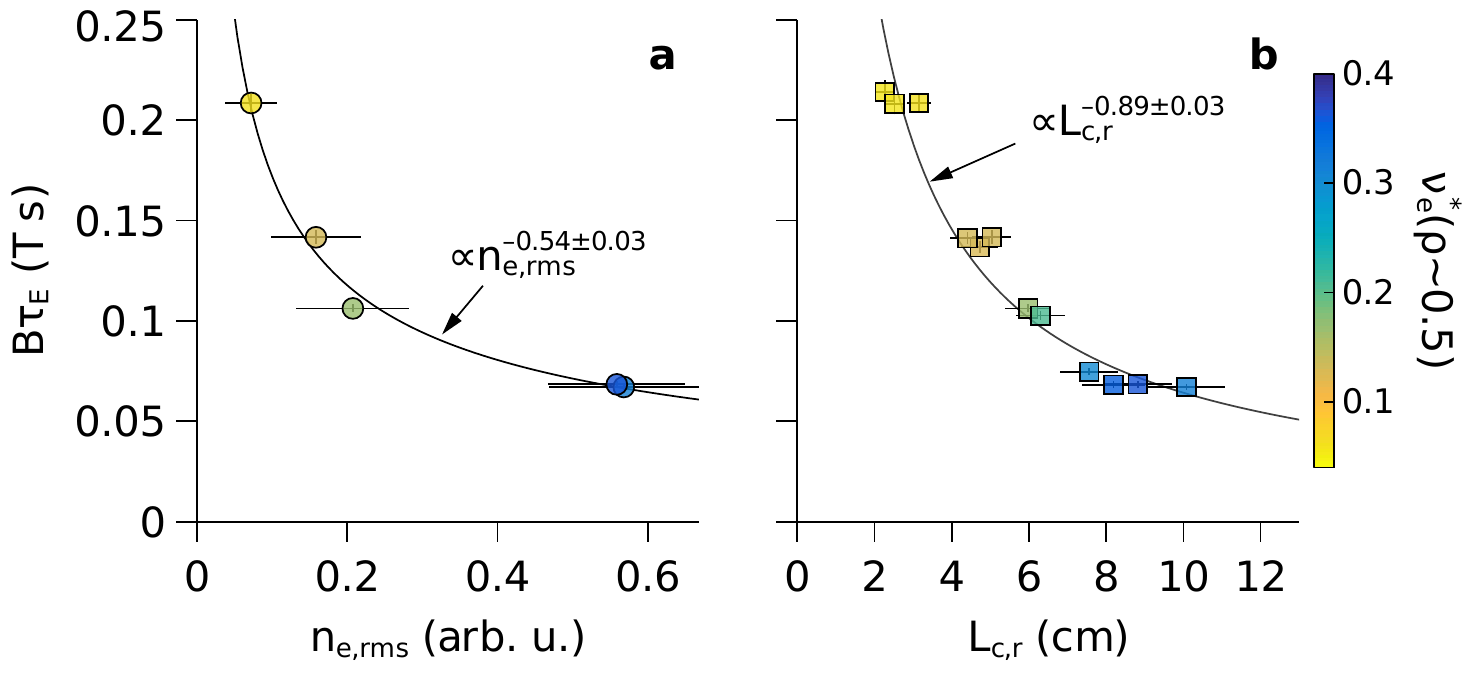}
  \caption{\label{fig:confinement_scaling} The normalized energy confinement time is plotted against (a) the amplitude ($k_{\theta}\rho_{s}\approx3$) and (b) the radial correlation length of the LRRC turbulent structures. The local collisionality is color-coded with darker markers corresponding to the higher collisionality shots.}
\end{figure}

It is found that the development of LRRC turbulent transport events is associated with the degradation of the energy confinement in these discharges.
As shown in \cref{fig:confinement_scaling}(a), the normalized energy confinement time decreases with increasing amplitude of LRRC turbulent transport events, i.e., $B\tau_{E}\propto|\tilde{n}_{e,\mathrm{LRRC}}|^{-0.54}$.
This scaling appears to be close to the collisionality dependence observed in previous DIII-D experiments, \cite{pettyScalingHeatTransport1999} i.e., $B\tau_{E}\propto{\nu^{*}}^{-0.56}$.
The normalized energy confinement time also decreases when the radial correlation length of LRRC turbulent structures increases (\cref{fig:confinement_scaling}(b)).
As noted in the previous section, the amplitude of ion-scale long-wavelength density fluctuations ($k_{\theta}\rho_{s}<1$) decreases slightly in the weak flow shear case (\cref{fig:bes_fluct}).
These findings indicate that while the LRRC turbulent events developed from high-$k_{\theta}$ is highly correlated to the confinement degradation, low-$k_{\theta}$ turbulence is unlikely responsible for the enhanced turbulent transport in the weak flow shear and high collisionality case.

\section{ETG mode as underlying linear instability\label{sec:linear_cgyro}}

\begin{figure}
  \centering
  \includegraphics[width=0.4\linewidth]{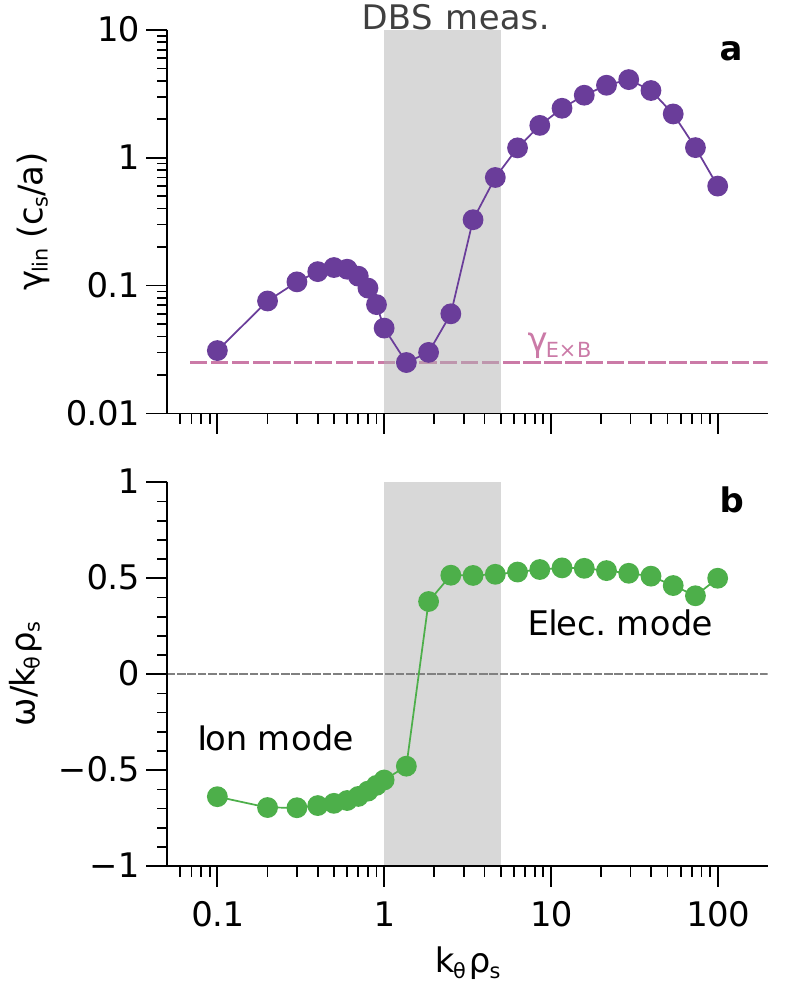}
  \caption{\label{fig:k_scan} \textsc{cgyro} calculations of the linear growth rate (a) and the phase velocity (b) for different wavenumbers at $\rho\approx0.6$ using experimental profiles. Positive velocity in (b) stands for the electron diamagnetic drift (EDD) direction. The shaded area indicates the wavenumber coverage of DBS measurements.}
\end{figure}

Linear gyro-kinetic simulations using the \textsc{cgyro} code\cite{candyHighaccuracyEulerianGyrokinetic2016} have been carried out to identify the underlying instabilities driving the LRRC turbulent transport events.
It can be seen in \cref{fig:k_scan}(a), for both low- and high-$k_{\theta}$ perturbations at $\rho\approx0.6$, the linear growth rates are greater than the local mean $E \times B$ flow shearing rate.
The low-$k_{\theta}$ perturbations ($k_{\theta}\rho_{s}<1$) propagate in the ion diamagnetic drift (IDD) direction, while the high-$k_{\theta}$ perturbations ($k_{\theta}\rho_{s}>2$) propagate in the electron diamagnetic drift (EDD) direction (\cref{fig:k_scan}(b)).
The DBS measurements performed in this study cover the intermediate-to-high wavenumber range ($1<k_{\theta}\rho_{s}<4$).
The perturbations in this wavenumber range are mostly unstable and propagate in the EDD direction, as indicated by the linear \textsc{cgyro} simulations.

\begin{figure}
  \includegraphics[width=0.4\linewidth]{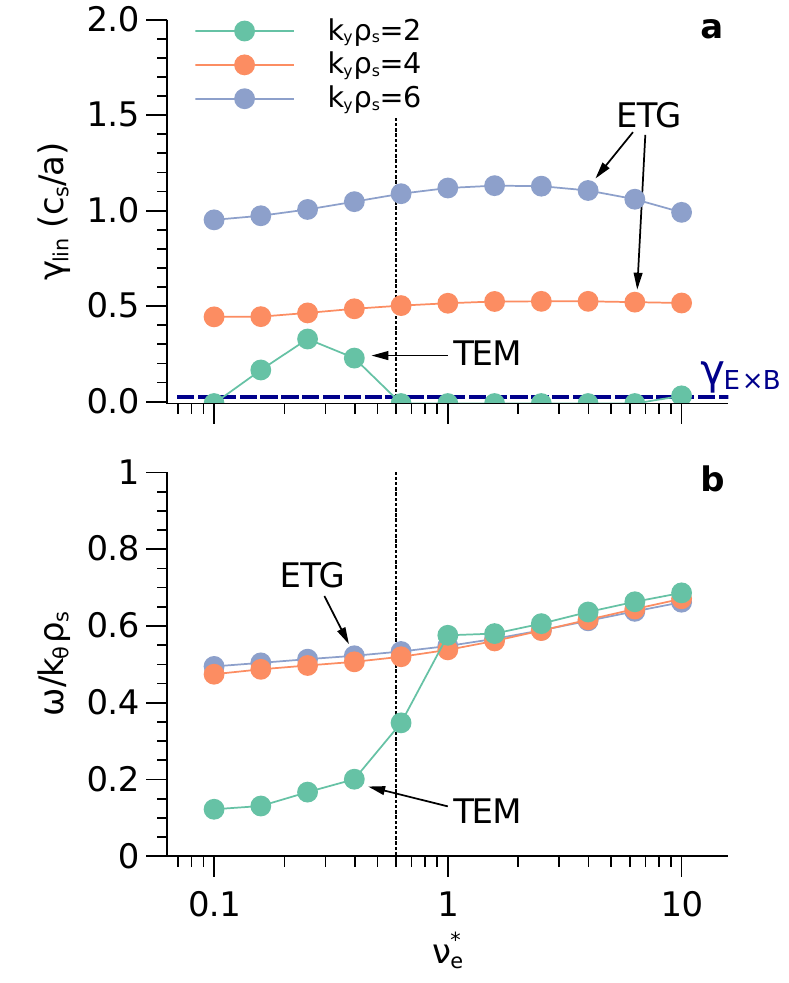}
  \caption{\label{fig:nu_scan} \textsc{cgyro} calculations of the linear growth rate (a) and the phase velocity (b) for different collisionality. Three different wavenumbers relevant to DBS measurements are used in this numerical scan. The experimental value of local collisionality at $\rho\approx0.6$ is indicated by the black dotted lines.}
\end{figure}

Two common micro-instabilities in this wavenumber range ($k_{\theta}\rho_{s}>2$) are the trapped electron mode (TEM) and the electron temperature gradient (ETG) mode.\cite{doyleChapterPlasmaConfinement2007}
Here, electromagnetic micro-instabilities are not considered because the magnetic fluctuations in corresponding wavenumber and frequency ranges have \emph{not} been detected in these discharges.
The TEM is predicted to be linearly stabilized in high-collisionality discharges.
A collisionality scan was performed in numerical simulations to verify this hypothesis for three different wavenumbers that are relevant to DBS measurements.
As shown in \cref{fig:nu_scan}(a), the linear growth rate of $k_{\theta}\rho_{s}=2$ mode is finite at low-$\nu^{*}$ but quickly reduces below the local mean flow shearing rate when the collisionality is raised to the experimental value.
It is also found that the poloidal phase velocity of $k_{\theta}\rho_{s}=2$ mode increases when the collisionality is raised (\cref{fig:nu_scan}(b)).
For higher-wavenumber modes, the linear growth rates and the poloidal phase velocity are not changed during the collisionality scan.
The reduced linear growth rate of the $k_{\theta}\rho_{s}=2$ mode is consistent with the suppression of TEMs at high collisionality.
Therefore, the TEM is not likely the underlying instability of the long-radial-range-correlation transport events.
The ETG mode, on the other hand, is not affected by the collisionality scan as shown in the numerical simulations (\cref{fig:nu_scan}(a)).
Previous nonlinear gyrokinetic simulations suggest ETG mode can lead to streamer-like transport events,\cite{dorlandElectronTemperatureGradient2000,jenkoElectronTemperatureGradient2000} which is in agreement with the observations in this study.
The ETG mode is thus a candidate instability responsible for long-radial-range-correlated transport events.

\section{Concluding remarks\label{sec:summary}}

In this study, we report on the observation of mesoscopic turbulent transport events in high-collisionality \emph{H}-mode plasmas with reduced mean $V_{E\times B}$ shear flow on the DIII-D tokamak.
These mesoscopic transport events develop from core turbulence with a sub-ion-gyro-radius scale ($1<k_{\theta}\rho_{s}<4$), and show a long-radial-range correlation (LRRC). 
In other words, these turbulent structures are radially elongated, and their envelopes span a wide range in the mid-radius region, leading to streamer-like transport events. 
The underlying turbulence is highly intermittent and shows a power spectrum of $S(k_{\perp}) \propto k_{\perp}^{-1}$ which is often associated with self-organized criticality.
The amplitude and the radial scale of the turbulent structures increase substantially once the mean flow shearing rate is decreased below the turbulent scattering rate. 
The observations summarized here provide evidence for the role of $V_{E\times B}$ shear flow in regulating the long-range correlated turbulent structures.
In addition, the normalized energy confinement time decreases with the increasing amplitude of the LRRC transport events, while the long-wavelength ions-scale turbulence remains approximately unchanged.
These findings suggest that the emergence of LRRC transport events may be a candidate explanation for the degradation of energy confinement time in high collisionality and weak flow shear shots.
Linear gyro-kinetic simulations show that the ETG mode is likely the underlying instability of the LRRC transport events.

%future work
It is worth noting that the reduced mean shear layer is associated with the high collisionality in this study.
The increased collisionality does not contribute to the enhanced electron thermal diffusivity as indicated by the \textsc{cgyro} simulations, since the TEM turbulence is stabilized by high collisionality while the ETG turbulence is not affected.
It is also suspected that the high collisionality can increase the collisional damping of zonal flows and hence increase the nonlinear saturation level of ambient turbulence.
However, the collisional damping of zonal flows is supposed to be applicable to both ion- and electron-scale turbulence. 
This hypothesis is not consistent with the observation of slightly decreased ion-scale density fluctuations at higher collisionality.
Nonlinear multiscale simulations would be helpful to distinguish the effects of collisionality and mean shear flows.
It thus calls for more detailed experimental studies using net-zero torque input, with either RF-only or balanced NBI heating.

\section*{Acknowledgments}

The authors greatly appreciate the effort and support of the entire DIII-D team in performing this experiment. 
We would like to acknowledge valuable discussions with M.~E.~Austin, N.~A.~Crocker, N.~T.~Howard, and S.~P.~Smith.
This material is based upon work supported by the U.S. Department of Energy, Office of Science, Office of Fusion Energy Sciences, using the DIII-D National Fusion Facility, a DOE Office of Science user facility, under Awards DE-FC02-04ER54698, DE-FG02-08ER54999, and DE-SC0019352. 

\section*{Disclaimer}
This report was prepared as an account of work sponsored by an agency of the United States Government. 
Neither the United States Government nor any agency thereof, nor any of their employees, makes any warranty, express or implied, or assumes any legal liability or responsibility for the accuracy, completeness, or usefulness of any information, apparatus, product, or process disclosed, or represents that its use would not infringe privately owned rights. 
Reference herein to any specific commercial product, process, or service by trade name, trademark, manufacturer, or otherwise does not necessarily constitute or imply its endorsement, recommendation, or favoring by the United States Government or any agency thereof.
The views and opinions of authors expressed herein do not necessarily state or reflect those of the United States Government or any agency thereof.

\bibliographystyle{apsrev4-2}
\bibliography{PoP_LRRC}

\end{document}